\RequirePackage{fix-cm}
\documentclass[twocolumn]{svjour3}          
\smartqed  
\usepackage{graphicx}
\begin{document}

\title{Elastic Instability of the Orthorhombic Antiferromagnetic Phase of 122-Pnictides Under Pressure\thanks{This work was supported in part by the DOE under grant DE-SC0014506}}




\author{Michael Widom        \and
        Khandker Quader 
}

\institute{M. Widom \at
              Dept of Physics, Carnegie-Mellon University, Pittsburgh, PA 15213, USA \\
              \email{widom@cmu.edu}           
           \and
           K. Quader \at
            Dept of Physics, Kent State University, Kent, OH 44242, USA  
}

\date{Received: date / Accepted: date}

\maketitle

\begin{abstract}
 $A$-122 pnictides ($A$ an alkaline earth element) possess striped antiferromagnetic states of 
 orthorhombic (OR)  symmetry and nonmagnetic states of tetragonal (T) symmetry.  Based on total energy calculations, we show that the $T=0$K pressure-driven transition from OR to T states
occurs at a pressure, $P_H$, where the tetragonal enthalpy drops below the orthorhombic.  The OR state remains metastable up to a higher pressure, $P_M>P_H$.  We find anomalies in magnetism and orthorhombicity as $P\rightarrow P_M$, and a trend towards elastic instability.


\end{abstract}

\section{Introduction}
\label{sec:intro}
The 122 pnictides, $AFe_2As_2$ ($A$ = Ca, Sr, Ba), display structural, magnetic, and superconducting phase transitions upon doping or applied pressure~\cite{Alireza09,Torikach09,Mani09,Saha11}. At ambient pressure, Ca-, Sr-, and Ba-122 compounds transition from the high-temperature tetragonal (T) phase to a low temperature orthorhombic (OR) phase, striped along the $a$-axis and antiferromagnetically (AFM) ordered along the $c$-axis, at N\'eel temperature $T_{N}$ $\sim$ 170K, 205 K, 140 K respectively~\cite{Torikach08,Krellner08,Huang08}. The T-OR transition temperature $T_{N}$ decreases with applied pressure.

At low-$T$, under hydrostatic pressure, the 122-pnictides exhibit rich universal behavior, {\it albeit} with interesting differences. While Ca-122 transitions from the OR-AFM phase to a collapsed tetragonal (cT) phase with strongly decreased $c$-axis at P $\sim$ 0.35 GPa~\cite{Canfield09}, Sr-122 and Ba-122 transition to a T phase with only slightly decreased $c$-axis, at P $\sim$ 4.4 GPa~\cite{ikeda12,kotegawa09} and 10.5 GPa respectively~\cite{Mittal11,Uhoya10}. These eventually transition to a cT phase at higher pressures.

Our previous work~\cite{Widom-Quader13}, based on first principles density functional theory calculations, showed that the pressure-driven $T = 0$K transition from OR-AFM to tetragonal (cT or T) can be understood as a 1$^{\rm st}$-order enthalpic transition that occurs at a pressure we denote $P_H$.  Subsequent work~\cite{Quader-Widom14,Quader-Widom15}, also based on first principles total energy calculations, proposed that the T - cT transition and anomalies in lattice parameters and elastic properties, observed experimentally at finite temperatures, can be interpreted as arising from proximity to $T = 0$K Lifshitz transitions, at pressures denoted as $P_L$, wherein pressure-driven changes in lattice constants alter the Fermi surface topology in these materials. 
Finally, our work revealed the existence of a metastable OR-AFM phase that remains mechanically stable until it transitions discontinuously to the tetragonal phase at a pressure we denoted $P_M$.

Here we explore $A$-122 pnictides in further detail, focusing on properties of the metastable OR-AFM phase.  We study the pressure dependence of relative enthalpy, magnetism, and lattice parameters, which all vanish discontinuously at $P_M$.  Additionally, elastic stability vanishes continuously as $P\rightarrow P_M$.

\section{Methods}
\label{sec:methods}

Our calculations follow methods outlined previously~\cite{Widom-Quader13,Quader-Widom14}, using projector-augmented wave density functional theory in the PBE generalized gradient approximation~\cite{PBE}, as implemented in VASP~\cite{Kresse96}.  
Owing to the previously reported~\cite{Quader-Widom14} sensitivity to calculational details, in the present work we take low Fermi surface smearing of $\sigma=0.05$ eV, with a high density 15x15x7 $k$-point grid for our 20-atom OR-AFM cells and matching 20-atom $\sqrt{2}\times\sqrt{2}$ tetragonal supercells.  We cut off our plane-wave basis at 350 eV which is 30\% greater than the default for Fe-based alloys.  Additional accuracy of forces is achieved though Fourier transform grids of full resolution, additional support grid for the evaluation of augmentation charges, and application of projectors in reciprocal space.  Self-consistency of electronic structure is achieved when subsequent iterations change the energy by less than $5\times 10^{-10}$eV.  Relaxations are considered converged when forces drop below 
$10^{-5}$eV/\AA~ and pressure matches its target to within $10^{-4}$GPa.  Relaxations are halted and restarted repeatedly to remove errors connected with volume-dependence of the basis sets.  To calculate elastic constants we employ central finite differences using a small step size of 0.010~\AA, because the stress-strain relation can be strongly nonlinear.

In order to locate the end point of metastability, $P_M$, we increment the pressure in steps from low to high, taking the relaxed structure at one pressure as the starting point for the next. 
Successively smaller steps are employed as the threshold pressure is approached.  Relaxation becomes intractable very close to $P_M$, because relaxation times diverge and error sensitivity grows. Hence we locate $P_M$ only to within 0.1 GPa.

\section{Results}
\label{sec:results}

To distinguish stable OR-AFM, metastable OR-AFM and T states, we monitor the enthalpy difference,
\begin{equation}
\label{eq:dH}
\Delta H=\Delta E+P\Delta V\equiv H_{\rm tetra}-H_{\rm ortho},
\end{equation}
the iron magnetic moment $M$, and orthorhombicity 
\begin{equation}
\label{eq:dab}
\delta_{ab}=\frac{a-b}{a+b}.
\end{equation}
We also monitor the elasticity tensor~\cite{Kittel8}, which for orthorhombic symmetry is
\begin{equation}
\label{eq:Cij}
{\bf C} = 
\left(
\begin{array}{cccccc}
C_{11} & C_{12} & C_{13} & 0 & 0 & 0 \\
C_{12} & C_{22} & C_{23} & 0 & 0 & 0 \\
C_{13} & C_{23} & C_{33} & 0 & 0 & 0 \\
0 & 0 & 0 & C_{44} & 0 & 0 \\
0 & 0 & 0 & 0 & C_{55} & 0 \\
0 & 0 & 0 & 0 & 0 & C_{66} \\
\end{array}
\right).
\end{equation}
Strains of Voigt type 1, 2, and 3 correspond to stretching, respectively, along the $x$, $y$ and $z$ axes, while type 4, 5, and 6 correspond to shears of types $yz, zx$ and $xy$.  Enthalpy, magnetism and orthorhombicity, and elasticity are discussed in turn in the following subsections.

\subsection{Enthalpy}
As described in detail in our previous work\cite{Quader-Widom14}, and shown in the top panels (a) of Figs. 1-3, each of the three $A$-122 compounds undergoes a 1$^{\rm st}$-order enthalpic transition from the OR-AFM phase to a non-magnetic tetragonal phase (cT or T) where $\Delta H$ crosses 0 at pressure $P_H$. However, a thermodynamically metastable OR-AFM state persists up to a higher pressure, $P_M>P_H$.  This metastable state has higher enthalpy than the competing nonmagnetic tetragonal state ({\em i.e.} $\Delta H<0$), but it retains mechanical stability.  Thus it constitutes a stable fixed point of conjugate gradient relaxation, considered as a dynamical system whose attractors are local energy minima.  Beyond $P_M$, this fixed point vanishes.  Relaxation carries the system towards an alternate fixed point, which turns out to be a nonmagnetic tetragonal state that is both thermodynamically and mechanically stable.  Hence, $\Delta H$ vanishes discontinuously for $P > P_M$, as can be seen in panel (a) of each figure.

Because the Lifshitz transition pressure $P_L$ is negative for Ca-122, while the metastable limit $P_M$ is positive, we find that $P_L<P_M$ for Ca-122.  Consequently, the metastable orthorhombic state relaxes to the collapsed tetragonal state with a low $c$-axis for $P>P_M$.  In contrast, for Sr-122 and Ba-122, we find that $P_M<P_L$, so the metastable orthorhombic state relaxes to the {\em un}collapsed tetragonal state with large $c$-axis for $P>P_M$, followed by a second transition (the Lifshitz transition~\cite{Quader-Widom14}) to the collapsed state with low $c$-axis at higher pressures $P_L>P_M$.

Values of $P_H$ and $P_M$ grow monotonically as we move down the alkaline earth column of the periodic table, from Ca through Sr to Ba.  However, the fractional interval of metastability $(P_M-P_H)/P_H$ diminishes from a high of 700\% for Ca-122 to a low of 5\% for Ba-122.  

\subsection{Magnetism and Orthorhombicity}
The tetragonal states (T and cT) lack magnetism and orthorhombicity.  Thus, $M$ and $\delta_{ab}$ vanish discontinuously at $P_M$ as can be seen in panel (b) of each figure.  They approach finite limiting values as $P$ approaches $P_M$ from below, while they vanish identically for all $P>P_M$.  Enlarging the data near the discontinuity (not shown) reveals nonanalyticity in which $M$ and $\delta_{ab}$ approach their limiting values with a singularity consistent with a square root power law.  The similar behavior of $M$ and $\delta_{ab}$ reflects the coupling of magnetism and lattice parameters.

\subsection{Elasticity}
\label{sec:elasticity}

Notice that ${\bf C}$ is block diagonal, with a symmetric 3x3 block for $i,j\le 3$, and three diagonal entries for $i=j\ge 4$.  Thus nine elastic constants are required to describe the elastic response in the OR-AFM state. Tetragonal symmetry can be regarded as a limiting case of orthorhombic symmetry for which orthorhombicity vanishes ($\delta_{ab}=0$) and ${\bf C}$ obeys the identities $C_{11}$=$C_{22}$, $C_{13}=C_{23}$ and $C_{44}=C_{55}$.  Values of $C_{ij}$ are plotted in panel (c) of each figure.

Elastic stability requires ${\bf C}$ to be positive definite, that is, its eigenvalues $\{\lambda_k\}$ must be positive.  This demands the matrix elements $C_{ij}$ obey a set of six constraints, two of which are nonlinear~\cite{Mouhat2014}.  As an alternative test of positivity, we plot the eigenvalues in panel (d) of each figure.  The eigenvalues are ordered so that $\lambda_k$ with $k=1-3$ denote the eigenvalues of the 3x3 block (in increasing order), while $k=4-6$ simply repeat the diagonal entries $C_{kk}$.  When elastic stability is violated a crystal will spontaneously deform through a combination of strains characterized by the eigenvector of the vanishing or negative eigenvalue.

All $C_{ij}$ exhibit anomalies as $P\rightarrow P_M$. The elastic constant $C_{33}$ in particular shows a strong dip for all $A$-122.  Low $C_{33}$ is in danger of violating the Born criterion derived from positivity of the determinant of the 3x3 block~\cite{Mouhat2014},
\begin{equation}
\label{eq:stable}
C_{33} >
\frac{C_{11}C_{23}^2+C_{22}C_{13}^2+2C_{12}C_{13}C_{23}}{C_{11}C_{22}-C_{12}^2},
\end{equation}
which is equivalent to the requirement $\lambda_1>0$. Near $P_M$ we find that $\lambda_1\approx 0$ to within our uncertainty that is limited by our difficulty in relaxation and our finite difference numerical differentiation.  $\lambda_1$ tends towards 0 with approximately a square root singularity.  The eigenvectors $v_1$ corresponding to eigenvalues $\lambda_1$ are superpositions of strains of types 1-3.  The associated deformation stretches the $c$-axis while shrinking the $a$- and $b$-axes (or {\em vice-versa}).  Orthorhombicity is evident in the differences between $C_{13}$ and $C_{23}$ which grow to around 50\% as $P\rightarrow P_M$, even while $C_{11}\approx C_{22}$ and $C_{44}\approx C_{55}$ are nearly consistent with tetragonality.

\section{Discussion}
\label{sec:discuss}

Relaxation of a crystal structure can be considered as a nonlinear dynamical system with dissipative dynamics~\cite{Chu2008}.  Coupling of the crystal atomic structure to the electronic band structure via the Kohn-Sham equations of density functional theory make the system effectively infinite dimensional.  Such dynamical systems typically exhibit multiple fixed points, and these fixed points can bifurcate as parameters are varied.  In the present case, we are confronted with the seeming disappearance of a stable fixed point corresponding to the OR-AFM state, as $P\rightarrow P_M$, which would seem to violate local conservation of Poincar\'e index~\cite{Guckenheimer83}.

The mystery could be resolved if the stable fixed point were to annihilate with an unstable fixed point, such as occurs in a saddle-node bifurcation.  The square-root singularities observed in magnetization, orthorhombicity, and elasticity are typical of such a bifurcation.  A concrete demonstration of this hypothesis would require observing the unstable fixed point.  Unfortunately this is not readily achievable, as our conjugate gradient dynamics is optimized to seek stable fixed points.

Considering that the numerical value of the threshold pressure $P_M$ is sensitive to Fermi surface smearing $\sigma$, it is interesting to ask if the loss of metastability could be caused by a Lifshitz transition (change in Fermi surface topology).  Indeed, we do find bands approaching to within $\sigma$ of the Fermi energy as $P\rightarrow P_M$ for all three pnictides.  However, the large unit cell with low symmetry and $s$, $p$, and $d$ elements inevitably creates many bands, leading to a proliferation of Lifshitz transitions, including several close to $P_M$, making it hard to assess the relevance. This raises the possibility that 
superconductivity which, in Sr- and Ba-122, occur in the vicinity of $P_M$, may be correlated with 
these Lifshitz transitions~\cite{Antonio10}.

A pronounced dip in $C_{33}$ is indicative of softening of associated phonon mode(s) as the system transforms
from OR-AFM to a T phase. This is reminiscent of Martensitic transitions in some of the A15 compounds~\cite{Testardi75},
except that here this occurs as a function of pressure at $T=0$. While the underlying mechanism for superconductivity in pnictides is believed to be magnetic in origin, we conjecture that phonon mode softening,
and strong magneto-elastic coupling that our calculations suggest, may be consequential to the 
pressure and temperature range over which superconductivity occurs.


\begin{figure}
\includegraphics[width=0.50\textwidth]{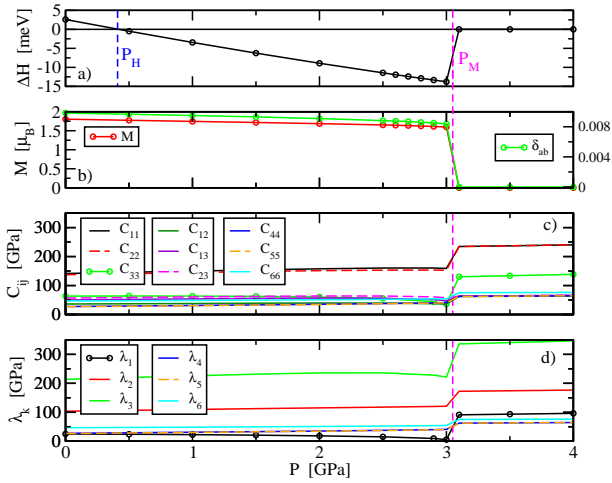}
\caption{Data for Ca-122. (a) Enthalpy difference $\Delta H$ (Eq.~(\ref{eq:dH}), units meV/atom), (b) magnetization $M$ (units $\mu_B$) and dimensionless orthorhombicity $\delta_{ab}$ (Eq.~(\ref{eq:dab})), (c) elastic constants $C_{ij}$ (units GPa), and (d) eigenvalues $\lambda_k$ (units GPa) all as functions of pressure $P$ (GPa).  Plotting symbols indicate pressures where calculations were performed.  Enthalpic transition at $P_H$ and metastability limit $P_M$ are marked.}
\label{fig:Ca}       
\end{figure}

\begin{figure}
\includegraphics[width=0.50\textwidth]{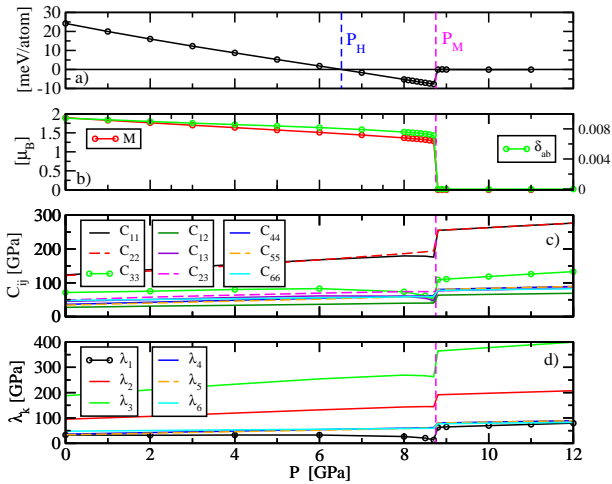}
\caption{Data for Sr-122. Other details as in Fig.~\ref{fig:Ca}.}
\label{fig:Sr}       
\end{figure}

\begin{figure}
\includegraphics[width=0.50\textwidth]{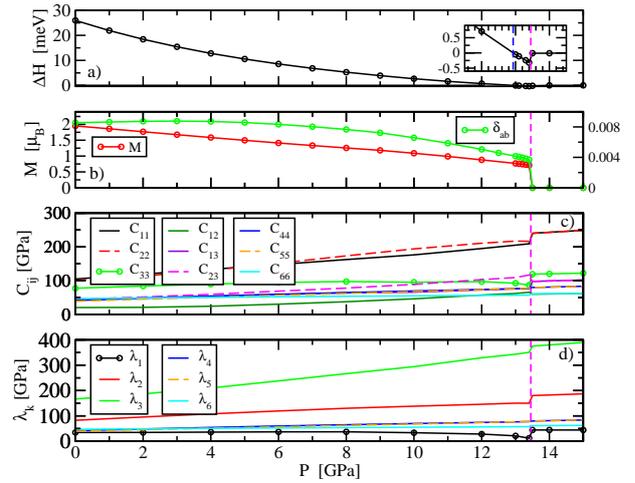}
\caption{Data for Ba-122. $P_H$ and $P_M$ marked on inset of (a).  Other details as in Fig.~\ref{fig:Ca}.}
\label{fig:Ba}       
\end{figure}
%


\bibliographystyle{./spphys}       
\bibliography{lifshitzw2}   

\end{document}